\documentstyle[aps,preprint]{revtex}
\begin{document}
\draft
\title{Gor'kov and Eliashberg Linear Response Theory: Rigorous
Derivation and Limits of Applicability}
\author{Ya.~M.~Blanter$^{a,b}$ and A.~D.~Mirlin$^{a,c}$}
\address{$^a$ Institut f\"ur Theorie der Kondensierten Materie,
Universit\"at Karlsruhe, 76128 Karlsruhe, Germany\\
$^b$ Department of Theoretical Physics, Moscow Institute for Steel and
Alloys, Leninskii Pr. 4, 117936 Moscow, Russia\\
$^c$ Petersburg Nuclear Physics Institute, 188350 Gatchina,
St. Petersburg, Russia}
\date{September 15, 1995}
\maketitle 
\tighten
\begin{abstract}
A rigorous microscopic calculation of the polarizability of disordered
mesoscopic particles within the grand canonical ensemble is given in  
terms of the supersymmetry method. The phenomenological result
of Gor'kov and Eliashberg is confirmed. Thus the underlying
assumptions of their method are justified. This encourages application
of RMT in the Gor'kov--Eliashberg style to more complicated situations.
\end{abstract}

Nearly 30 years ago Gor'kov and Eliashberg (GE) in the pioneering paper
\cite{GE} applied the random matrix theory (RMT) and the concept of level
correlation to the condensed matter physics. They studied the
polarizability of a small (mesoscopic) disordered metallic
particle. The dipole 
moment of the particle in a weak electric field $\mbox{\bf E}(t)$ is
given by the Kubo formula:
\begin{equation} \label{dip}
\langle d_i (t) \rangle = i\int_{-\infty}^{t} \left\langle
\left\langle [d_i(t),d_j(t')] \right\rangle_T \right\rangle E_j(t') dt'
\end{equation}
Here the square brackets stand for the commutator while the angular
ones denote the impurity $\langle \dots \rangle$ and the
thermodynamic $\langle \dots \rangle_T$ averages. Introducing a set
of eigenstates $\vert k \rangle$ and eigenvalues $\epsilon_k$ of the
Hamiltonian in a particular disorder configuration, one can rewrite
Eq. (\ref{dip}) as
\begin{equation} \label{dip1}
\langle d_i(\omega) \rangle = e^2 E_j(\omega) \left\langle \sum_{k,l}
\frac{n_F(\epsilon_l) - n_F(\epsilon_k)}{\epsilon_k - \epsilon_l +
\omega + i\delta} (r_i)_{kl} (r_j)_{lk} \right\rangle
\end{equation} 
Here $(r_i)_{kl}$ is the matrix element of the coordinate:
$(r_i)_{kl} = \langle k \vert \hat r_i \vert l \rangle$, and $\hbar =
1$. The consequent analysis of Eq. (\ref{dip1}) by GE is based on the
following assumptions:
\begin{description}

\item[(i)] The product of the matrix elements in Eq.(\ref{dip1}) can
be averaged independently of the remaining part of the expression;

\item[(ii)] This averaging can be performed quasiclassically;

\item[(iii)] The remaining summation over the energy levels can be
performed with the use of the two-level correlation function known
from the random matrix theory \cite{matr}.
\end{description}

This semi-phenomenological calculation inspired a number of attempts to
improve it. First, Str\"assler, Rice, and Wyder \cite{Wyder} have
pointed out that GE calculated in fact the response to the local
electric field rather than to the external one. We will return to this
point later. Then, Devaty and Sievers \cite{Devaty} corrected some
minor mistakes contained in the GE paper. Shklovskii \cite{Shkl}, and
later Sivan and Imry 
\cite{SivanImry} considered the modification of the GE result for the
canonical ensemble. Efetov \cite{Efetov} derived the two-point
level correlator for a disordered granule by means of
the supersymmetry method and showed that it has exactly the
same form as in RMT, thus having justified the assumption (iii) of
GE. Finally, Frahm, M\"uhlschlegel, and N\'emeth 
\cite{FMN} attacked the assumption (ii) and tried to improve the GE result by
introduction of frequency-dependent averaged matrix elements. However,
for small frequencies $\omega \ll E_c$ with $E_c = D/L^2$ being the
Thouless energy (here $D$ and $L$ are the diffusion coefficient and
the size of the system respectively) this dependence proved to be
inessential and they have just reproduced the GE result for the matrix
elements.   

Below we revisit this problem completely. We restrict ourselves to the
case of diffusive mesoscopic system ($l \ll L$, with $l$ being the
mean free path) and grand canonical ensemble. For simplicity we
consider only the case of Gaussian Unitary Ensemble (GUE). 
We present a {\em microscopic, well-controlled} 
calculation, based on the supersymmetry method, that does not involve
GE assumptions (i), (ii), and (iii). We demonstrate that the GE result is
correct (to some extent, which we also analyze), and hence the
assumptions (i) and (ii) are justified. Consequently we are able to
calculate responses of rather general form, and we discuss the
conditions of applicability of the GE method. 

For our purpose the formula (\ref{dip}) may be conveniently rewritten
in terms of the Matsubara Green's functions. In the Fourier representation
we have $d_i (\omega) = \alpha_{ij} (\omega) E_j(\omega)$,
$\alpha_{ij} = \alpha \delta_{ij}$,
and the polarizability $\alpha(\omega)$ is:
\begin{equation} \label{alpha1}
\alpha(i\omega_n) = - \frac{e^2}{3} T\sum_{\epsilon_m} \int \mbox{\bf r}_1
\mbox{\bf r}_2 d\mbox{\bf r}_1
d\mbox{\bf r}_2  \langle G(\mbox{\bf r}_1, \mbox{\bf r}_2, 
i\epsilon_m + i\omega_n) G(\mbox{\bf r}_2, \mbox{\bf r}_1, i\epsilon_m) 
\rangle
\end{equation}
The integrand contains a combination $\mbox{\bf r}_1 \mbox{\bf r}_2$, which
is not translationally invariant. This can be cured with the use
of the identity
\begin{equation} \label{id}
T\sum_{\epsilon_m} \int d\mbox{\bf r}_1
\langle G(\mbox{\bf r}_1, \mbox{\bf r}_2, \epsilon_m + \omega_n)
G(\mbox{\bf r}_2, \mbox{\bf r}_1, \epsilon_m) \rangle = 0,
\end{equation}
which enables us to replace this combination by a translationally
invariant one $-(\mbox{\bf r}_1 -\mbox{\bf r}_2)^2/2$. 
After the analytical continuation $i\omega_n \to \omega$ is performed,
Eq. (\ref{alpha1}) is expressed through the advanced and retarded Green's
functions as follows:  
\begin{eqnarray} \label{alpha}
\alpha(\omega) & = & \frac{ie^2}{12\pi} \int d\epsilon d\mbox{\bf r}_1
d\mbox{\bf r}_2 \left(\mbox{\bf r}_1 - \mbox{\bf r}_2 \right)^2 \left
\{ n_F(\epsilon) \langle G^{R} 
(\mbox{\bf r}_1, \mbox{\bf r}_2, \epsilon + \omega) G^{R} (\mbox{\bf r}_2,
\mbox{\bf r}_1, \epsilon) \rangle - \right. \nonumber \\
& - & \left. n_F(\epsilon + \omega) \langle G^{A}
(\mbox{\bf r}_1, \mbox{\bf r}_2, \epsilon + \omega) G^{A} (\mbox{\bf r}_2,
\mbox{\bf r}_1, \epsilon) \rangle + \right. \nonumber \\
&+& \left. [n_F(\epsilon + \omega) - n_F(\epsilon)]  \langle G^{R}
(\mbox{\bf r}_1, \mbox{\bf r}_2, \epsilon + \omega) G^{A} (\mbox{\bf r}_2,
\mbox{\bf r}_1 \epsilon) \rangle \right\} 
\end{eqnarray} 

The products $ \langle G^RG^R \rangle$ and $\langle G^AG^A \rangle$
can be calculated in the usual impurity perturbation theory
(see, e.g. \cite{AA}). Direct calculation gives: 
$$\int d\epsilon \left \{ n_F(\epsilon) \langle G^{R}
(\mbox{\bf r}_1, \mbox{\bf r}_2, \epsilon + \omega) G^{R} (\mbox{\bf r}_2,
\mbox{\bf r}_1, \epsilon) \rangle - n_F(\epsilon+\omega) \langle G^{A}
(\mbox{\bf r}_1, \mbox{\bf r}_2, \epsilon + \omega) G^{A} (\mbox{\bf r}_2,
\mbox{\bf r}_1, \epsilon) \rangle \right\} \approx $$ 
\begin{equation} \label{RRAA}
\approx \frac{i\nu}{4r^4}\exp(-\frac{r}{l}) \left[\sin 2p_Fr - 2p_Fr\cos
2p_Fr \right] + \omega \left( \frac{\pi\nu}{p_Fr} \right )^2
\exp(-\frac{r}{l}) \cos(2p_Fr)
\end{equation}
Here $r = \vert \mbox{\bf r}_1 - \mbox{\bf r}_2 \vert$, and $\nu$ is
the density of states.

At the same time the perturbation theory is not valid for the $\langle
G^RG^A \rangle$ term, and we involve the supersymmetry technique
\cite{Efetov,Verb} to calculate it. After the standard manipulations we
get 
\begin{equation} \label{Q1}
\langle G^RG^A \rangle (\mbox{\bf r}_1, \mbox{\bf r}_2, \epsilon,
\omega) = -\int DQ 
\left\{ g_{bb}^{11} (\mbox{\bf r}_1, \mbox{\bf r}_2) g_{bb}^{22}
(\mbox{\bf r}_2, \mbox{\bf r}_1) + g_{bb}^{12} (\mbox{\bf r}_1, \mbox{\bf
r}_1) g_{bb}^{21} (\mbox{\bf r}_2, \mbox{\bf r}_2) \right\} e^{-F[Q]}   
\end{equation}
Here $F[Q]$ is the action of the supermatrix sigma model:
\begin{equation} \label{F}
F[Q] = \frac{\pi\nu}{4} \int d\mbox{\bf r} \ \mbox{Str} [D(\nabla Q)^2 +
2i(\omega + i\delta) \Lambda Q],
\end{equation}
$Q = T^{-1}\Lambda T$ is a 4$\times$4 supermatrix, $\Lambda =
\mbox{diag} (1,1,-1,-1)$, and $T$ belongs to the supercoset space
$U(1,1 \vert 2)$. The symbol $\mbox{Str}$ denotes the supertrace defined as
$\mbox{Str} A = A_{bb}^{11} - A_{ff}^{11} + A_{bb}^{22} - A_{ff}^{22}$. The 
upper matrix indices correspond to the retarded-advanced
decomposition, while the lower indices denote the boson-fermion
one. The Green's function $g$ in Eq. (\ref{Q1}) is the solution to
the matrix equation:
\begin{equation} \label{Green}
\left[ -i(\epsilon + \frac{\omega}{2} - \hat H_0) - \frac{i}{2}(\omega
+ i\delta)\Lambda + Q/2\tau \right] g(\mbox{\bf r}_1, \mbox{\bf r}_2) =
\delta (\mbox{\bf r}_1 - \mbox{\bf r}_2)
\end{equation}
We consider now two terms in Eq. (\ref{Q1}) separately. The first one
yields (cf. \cite{PAEI}):
\begin{eqnarray} \label{RA}
& & -\langle g^{11}_{bb} (\mbox{\bf r}_1, \mbox{\bf r}_2) g^{22}_{bb}
(\mbox{\bf r}_2, \mbox{\bf r}_1) \rangle_F \approx \nonumber \\
& & \approx \left\{ \begin{array}{lr}
[\mbox{Re}\  G^R(\mbox{\bf r}_1, \mbox{\bf r}_2)]^2 - [\mbox{Im}\
G^R(\mbox{\bf r}_1, 
\mbox{\bf r}_2)]^2 \langle Q^{11}_{bb} (\mbox{\bf r}_1)
Q^{22}_{bb}(\mbox{\bf r}_1) \rangle_F,&\ \ \ r \ll l \\ 
0,& r \gg l
\end{array}
\right.
\end{eqnarray}
Here $G^R$ is the impurity-averaged retarded Green's functions: 
$$G^R(\mbox{\bf r}_1, \mbox{\bf r}_2) = G^{A*}(\mbox{\bf r}_1, \mbox{\bf
r}_2) = \frac{\pi\nu}{p_F r} \exp[ip_Fr - r/2l]$$
and $\langle \dots \rangle_F$ denotes the averaging with the sigma
model action $F[Q]$ (Eq.(\ref{F})). 

For relatively low frequencies $\omega \ll E_c$ the sigma-model
correlators for a closed metallic systems can be approximately
calculated within the so-called zero-mode approximation \cite{Efetov}.
This means that only the spatially constant configurations of the
field $Q(\mbox{\bf r})$ are taken into account, so that the
functional integral over $DQ(\mbox{\bf r})$ is reduced to an integral
over a single matrix $Q$. The latter can be calculated with the use of
the technique developed in Refs. \cite{Efetov,Verb,Zirn}. In
particular, the result for the correlator entering rhs of
Eq.(\ref{RA}) reads as
\begin{equation} \label{1122}
\langle Q_{bb}^{11}Q_{bb}^{22} \rangle = -1 -
\frac{2i\Delta^2}{\pi^2\omega^2}\exp \frac{\pi i \omega}{\Delta} \sin
\frac{\pi \omega}{\Delta}
\end{equation}
with $\Delta = (\nu V)^{-1}$ being the mean level spacing.

We will show below that the leading contribution to the polarizability
is given by the second term in Eq.(\ref{Q1}) due to the short-range
nature of the first one, Eq.(\ref{RA}). For this reason, we do not
try to calculate the first term with a better precision and turn to
the second one. It can be rewritten as 
\begin{equation} \label{1221}
-\langle g^{12}_{bb} (\mbox{\bf r}_1, \mbox{\bf r}_1) g^{21}_{bb}
(\mbox{\bf r}_2, \mbox{\bf r}_2) \rangle_F = -(\pi\nu)^2 \langle
Q_{bb}^{12} (\mbox{\bf r}_1) Q_{bb}^{21} (\mbox{\bf r}_2) \rangle_F
\end{equation} 

In the zero-mode approximation we obtain
$$ \langle Q_{bb}^{12} (\mbox{\bf r}_1) Q_{bb}^{21} (\mbox{\bf r}_2)
\rangle_F \approx - \frac{2i\Delta}{\pi\omega}$$
and eventually,
$$\alpha(\omega) = \frac{e^2}{6} \Delta \nu^2 \int d\mbox{\bf r}_1
d\mbox{\bf r}_2 (\mbox{\bf r}_1 -\mbox{\bf r}_2)^2$$
i.e. the result is frequency independent and real. Thus, in order to
obtain the imaginary part of $\alpha(\omega)$ (which determines the
conductivity) and to study the frequency dependence of the real part
we have to go beyond the zero-mode approximation in treatment of the
correlator $\langle Q_{bb}^{12}(\mbox{\bf r}_1)Q_{bb}^{21}(\mbox{\bf
r}_2) \rangle_F$. For this purpose, we use the method developed in
Refs. \cite{KM,MF}. The idea of this method is to decompose the matrix
$Q(\mbox{\bf r})$ into the constant part $Q_0$ (zero mode) and the 
spatially dependent contribution (non-zero modes); then the latter
should be integrated out. This step is
performed perturbatively, with the parameter being $g^{-1} =
\Delta/E_c \ll 1$. The next step is to calculate non-perturbatively
the integral over the matrix $Q_0$. 

Prior to the calculation it is convenient to rewrite
$\langle Q_{bb}^{12}Q_{bb}^{21} \rangle$ in terms of a supertrace:
\begin{equation} \label{STR}
\langle Q_{bb}^{12} (\mbox{\bf r}_1) Q_{bb}^{21} (\mbox{\bf r}_2)
\rangle = \frac{1}{4} 
\int DQ \ \mbox{Str} [\hat a Q(\mbox{\bf r}_1) \hat b Q(\mbox{\bf r}_2)] 
e^{-F[Q]} 
\end{equation}
Here we have introduced matrices 
$\hat a  = \mbox{diag}(1, -1, 0, 0)$ and $\hat b  = \mbox{diag}(0, 0,
1, -1)$. The matrix $Q$ can be parameterized
as follows:
$$Q(\mbox{\bf r}) = T_0^{-1}\Lambda \left( 1+ \frac{W(\mbox{\bf r})}{2}
\right) \left( 1- \frac{W(\mbox{\bf r})}{2} \right)^{-1} T_0 =$$ 
$$= T_0^{-1}\Lambda \left[1 + W + \frac{1}{2}
W^2 + \frac{1}{4}W^3 + \frac{1}{8}W^4 + \dots\right]T_0$$ 
where the matrix $W$ is block off-diagonal and does not contain a
contribution from the zero-wave-vector mode: $\int W(\mbox{\bf r})
d\mbox{\bf r} = 0$. Now one
has to expand the exponent and the pre-exponential factor in powers of
$W$; then the integration over the fast modes can be easily performed
using the Wick's theorem in the same way as it was suggested in
Ref. \cite{KM}. Restricting ourselves to the leading correction to the
zero-mode approximation (terms quadratic in $W$), we find:
\begin{equation} \label{Q2} 
\langle Q_{bb}^{12} (\mbox{\bf r}_1) Q_{bb}^{21} (\mbox{\bf r}_2)
\rangle = \int dQ_0 
\left\{ \mbox{Str}(\hat a Q_0 \hat b Q_0) - 
\frac{\Pi(\mbox{\bf r}_1,\mbox{\bf
r}_2)}{4} \left[ \mbox{Str}(\hat a Q_0) 
\mbox{Str} (\hat b Q_0) - 4 \right] \right\} \times 
\end{equation}
$$\times \exp \left[ \frac{\pi i}{2
\Delta} (\omega + i\delta) \mbox{Str} (\Lambda Q_0) \right]$$
The diffusion propagator $\Pi$ is the solution to the diffusion
equation
\begin{equation} \label{diff} 
\frac{\partial\Pi(\mbox{\bf r}_1,\mbox{\bf r}_2)}{\partial t} - D
\nabla^2\Pi(\mbox{\bf r}_1,\mbox{\bf r}_2) = (\pi \nu)^{-1}
\delta(\mbox{\bf r}_1 - \mbox{\bf r}_2)   
\end{equation}
with appropriate boundary conditions. In particular, if the system is
a cube with a size $L$, we obtain
\begin{equation} \label{pi}
\Pi(\mbox{\bf r}_1,\mbox{\bf r}_2) = \sum_{\mbox{\bf q}}
\frac{1}{\pi \nu V}\frac{1}{Dq^2 - i\omega} \cos(q_xx_1) \cos(q_yy_1)
\cos(q_zz_1) \cos(q_xx_2) \cos(q_yy_2) \cos(q_zz_2);
\end{equation}
$$\mbox{\bf q} = \frac{\pi}{L}(n_x,n_y,n_z), \ \ \ n_i = 0, \pm 1, \pm
2, \dots, \ \ \ \sum n_i^2 > 0$$  
Since we assume $\omega \ll E_c$ and the mode $\mbox{\bf q}=0$ does
not contribute to the sum (\ref{pi}), one can neglect $\omega$ in the
denominator. 

Now the integral over $Q_0$ can be calculated exactly, yielding:
\begin{equation} \label{Q3}
\langle Q_{bb}^{12} (\mbox{\bf r}_1) Q_{bb}^{21} (\mbox{\bf r}_2) \rangle = 
-2\left\{ \frac{i\Delta}{\pi\omega} + \left[ 1 + i\left( 
\frac{\Delta}{\pi\omega} \right)^2 \exp \left( \frac{\pi i
\omega}{\Delta} \right) \sin \left(\frac{\pi \omega}{\Delta} \right)
\right] \Pi(\mbox{\bf r}_1, \mbox{\bf r}_2) \right\}
\end{equation}

In order to get the final expression for the polarizability we
should combine Eqs. (\ref{alpha}), (\ref{RRAA}), (\ref{RA}) and
(\ref{Q3}). First we stress that the contributions of $\langle
G^RG^R\rangle$, $\langle G^AG^A \rangle$, and of the term (\ref{RA})
are short-ranged (localized on distances $r = \vert \mbox{\bf r}_1 -
\mbox{\bf r}_2 \vert$ of order of the mean free path $l$), while that
of Eq. (\ref{Q3}) is long-ranged. It is easy to check that the
contribution of the short-range terms to the frequency-dependent part
of $\alpha(\omega)$ is suppressed for arbitrary frequency by the
factor $(l/L)^2 \ll 1$. Thus, the contribution of the $\langle
Q_{bb}^{12} Q_{bb}^{21} \rangle$ term, Eq. (\ref{Q3}), dominates, and
we obtain 
\begin{equation} \label{fin}
\alpha(\omega) = -\frac{ie^2\omega}{6\pi} (\pi\nu)^2 \left\{
\frac{i\Delta}{\pi \omega} \int (\mbox{\bf r}_1 - \mbox{\bf r}_2)^2 
d\mbox{\bf r}_1 d\mbox{\bf r}_2 + \right.
\end{equation}
$$\left. + \left[ 1 + i\left(
\frac{\Delta}{\pi\omega} \right)^2 \exp \left( \frac{\pi i
\omega}{\Delta} \right) \sin \frac{\pi \omega}{\Delta} \right]
\int (\mbox{\bf r}_1 - \mbox{\bf r}_2)^2 \Pi(\mbox{\bf r}_1, \mbox{\bf r}_2) 
d\mbox{\bf r}_1 d\mbox{\bf r}_2 \right\}$$ 
This expression is valid for any closed or nearly closed \cite{Zirn1}
diffusive system irrespective to the geometry. For the
specified cubic geometry we have
\begin{equation} \label{finn}
\mbox{Re} \ \alpha(\omega) = \frac{e^2\nu L^5}{6} -
\frac{e^2\beta\omega}{\pi^6D} L^7\nu
\left(\frac{\Delta}{\pi\omega}\right)^2 \sin\frac{2\pi\omega}{\Delta} 
\end{equation}
$$\mbox{Im} \ \alpha(\omega) = \frac{2e^2\beta\omega}{\pi^6D} L^7\nu
\left[1 - \left(\frac{\Delta}{\pi\omega}\right)^2 \sin^2
\frac{\pi\omega}{\Delta}\right]$$ 
with
$$\beta = \sum_{k=0}^{\infty} \frac{1}{(2k+1)^6} = \frac{\pi^6}{960}$$ 

The result (\ref{finn}) is in agreement with the GE one (up to the numerical
coefficient, caused by the difference in geometries). 
Moreover, the general expression (\ref{fin}) is equivalent to the
corresponding GE expression. This becomes clear when we realize that the
time-integrated conditional probability introduced by GE ($\int dt
W_{\mbox{\bf r}}(\mbox{\bf r}',t)$ in their notations) is essentially
the same as our diffusion propagator $\Pi(\mbox{\bf r},\mbox{\bf
r}')$. This equivalence confirms the validity of the assumptions 
(i), (ii) and (iii), introduced phenomenologically by GE. It is
instructive to reformulate these assumptions in terms of the exact
single-particle states $\psi_k(\mbox{\bf r}) \equiv \vert k
\rangle$. Then the assumption (i) means that the amplitudes of these
states and the corresponding eigenvalues may be averaged separately,
while the assumption (ii) implies the following relation:
\begin{equation} \label{states}
\langle \psi^*_k(\mbox{\bf r}) \psi_l(\mbox{\bf r}) \psi_k(\mbox{\bf
r}') \psi^*_l(\mbox{\bf r}') \rangle = V^{-2} \Pi(\mbox{\bf
r},\mbox{\bf r}'), \ \ \ k \ne l 
\end{equation}
As is seen from Eq.(\ref{states}), the diffusion leads to the
long-range correlations of the eigenfunctions. This effect can not be
obtained in the zero-mode approximation (Gaussian ensemble).

To what extent are the assumptions (i) and (ii) exact? In order to
answer this question, we have to note that in the same manner as it
was done above for the polarizability we are able to evaluate any
quantity given by  
\begin{equation} \label{just}
\lambda(\omega) = T\sum_{\epsilon_m} \int \hat f(\mbox{\bf
r}_1,\mbox{\bf r}_2) d\mbox{\bf r}_1
d\mbox{\bf r}_2 \langle G(\mbox{\bf r}_1, \mbox{\bf r}_2, \epsilon_m +
\omega_n) G(\mbox{\bf r}_2, \mbox{\bf r}_1, \epsilon_m) \rangle
\end{equation}
with $\hat f$ being {\em arbitrary} operator. Alternatively,
one can treat Eq. (\ref{just}) by GE method. As we have seen, in a
particular case $\hat f =(\mbox{\bf r}_1 - \mbox{\bf r}_2)^2$ the results
coincide. It is straightforward to see, however, that for $\hat f
= 1$ the exact calculation gives zero due to the identity (\ref{id}),
while the RMT calculation in the GE manner would yield a finite
result. The reason is that RMT is able to take into account only the
long-ranged part of $\langle G^RG^A \rangle$, while other
(short-ranged) correlators are 
always lost. The polarizability is determined by the long-distance
behavior of the integrand, and so the RMT result is correct. However, in
cases when the short-distance behavior is important, the GE method
fails. Summing up, we can say that the GE-style calculation gives the
correct result only in the case when the dominant contribution to the
integral (\ref{just}) comes from the large distance range $r = \vert
\mbox{\bf r}_1 - \mbox{\bf r}_2 \vert \gg l$, which takes place when
$\hat f$ grows with $r$.   

The following comment is appropriate here. Following GE, we have
calculated the response to the uniform electric field and neglected
effects of screening. The true polarizability is however a response to
the external field rather than to the local one. Effect of the
screening on the polarizability was thoroughly studied in
Refs. \cite{Wyder,SivanImry}. Besides, it was noted in
Ref. \cite{Wyder} that non-uniformity of the local field may be
important. We should mention in this context that the ``local''
polarizability $\chi$, defined as the response to the non-uniform
electric field $\mbox{\bf E}(\mbox{\bf r})$,
$$\mbox{\bf d}(\mbox{\bf r}) = \int \chi (\mbox{\bf r}, \mbox{\bf r}')
\mbox{\bf E}(\mbox{\bf r}') d\mbox{\bf r}'$$
is given by the expression (\ref{fin}) without the integration over
coordinates. This quantity may be directly used for the calculation of
the true polarizability.

To conclude, we have provided a {\em rigorous microscopic calculation}
of the linear response expression for the frequency-dependent
polarizability of a disordered metallic 
particle. Our result is in agreement with that of Gor'kov and
Eliashberg \cite{GE}, thus justifying the underlying assumptions of
their derivation. This encourages application of RMT {\em \`a la}
Gor'kov--Eliashberg to more complicated situations. As a very
important example, we mention the first steps in non-perturbative
treatment of interacting systems \cite{Blanter,Zhou}. Since the
higher-order level correlation functions are involved in that case, an
attempt to employ the supersymmetry method meets enormous
complications. Therefore, RMT (in spirit of GE calculations) is the
only available approach to handle these problems. We should stress
however on the basis of the above analysis that the assumptions
analogous to (i), (ii), (iii) by GE are not always valid, and in any
particular case require a careful investigation.

The work was supported by the Alexander von Humboldt Foundation
(Y.~M.~B.) and SFB195 der Deutschen Forschungsgemeinschaft
(A.~D.~M.).

\end{document}